\documentclass [12pt]{article}
\def\maj#1{\ifmmode\mbox{\usefont{U}{msb}{m}{n}#1}\else{\usefont{U}{msb}{m}{n}#1}\fi}
\def\v#1{\mathbf{#1}}
\def\t#1{\tilde{#1}}
\usepackage{graphics,epsfig,graphicx}
\usepackage{color}
\makeatletter
\@addtoreset{equation}{section}
\makeatother
\renewcommand{\theequation}{\arabic{section}.\arabic{equation}}

\begin{document}

\title{\textbf{Third order susceptibility: general formalism for photoinduced current density in semiconductors}}
\author{M. Combescot and O. Betbeder-Matibet
\\ \small{\textit{Institut des NanoSciences de Paris,}}\\
\small{\textit{Universit\'e Pierre et Marie Curie, 
CNRS,}}\\
\small{\textit{Campus Boucicaut, 140 rue de Lourmel, 75015 Paris}}}
\date{}
\maketitle

\begin{abstract}
This paper contains a detailed derivation of the photoinduced current density at third order in the coupling between a semiconductor and a multifrequency photon field, starting from its standard textbook expression as a third order time integral of a triple commutator. Due to a major intrinsic problem linked to this triple commutator, such a derivation has been made possible quite recently only, thanks to the tools developed in the composite-boson many-body theory we have just constructed. The photoinduced current density is shown to ultimately read in a compact form, in terms of the Pauli and Coulomb scatterings for exciton-exciton interactions introduced in this theory. Representation in Shiva diagrams is also given to better grasp the physics of the various contributions.
\end{abstract}

PACS number: 71.35.-y

\newpage

\section{Introduction}

Nonlinear effects induced by unabsorbed photons are known to be of high technological interest due to the timescale on which ultrashort laser pulses allow them to operate. A quite usual way to describe these optical effects is through non-linear susceptibilities [1-17]. Since photons do not interact directly, these nonlinearities can only come from interactions between the matter excitations to which these unabsorbed photons are coupled. In the case of semiconductors, these matter excitations can be conveniently seen as virtual excitons. Interactions between excitons have two quite different origins: Coulomb interaction between the carriers of these excitons, but also and mainly Pauli exclusion between these carriers. This Pauli exclusion remained for decades quite difficult to handle properly. This is why excitons have been commonly treated as elementary bosons with effective scatterings dressed by a certain amount of carrier exchanges [18,19]. 

Over the last few years, we have developed a many-body theory for composite bosons made of two fermions, like the excitons, which, as a main goal, has the exact treatment of fermion exchanges between  composite bosons [20,21]. This  theory, in a natural way, generates ``Pauli scatterings'' which describe carrier exchanges between two excitons in the absence of carrier interactions. By combining these 2$\times$2 scatterings, it is actually possible to describe all carrier exchanges which can exist between $N$ excitons, these multiple exchanges being nicely visualized through the so-called ``Shiva diagrams'' [21,22]. Since these Pauli scatterings are dimensionless by construction, they in fact control all optical nonlinearities when the detuning increases, due to a bare dimensional argument: Coulomb scatterings associated to Coulomb interaction between carriers are energylike quantities, so that when they appear, they must have energylike denominators which can only be photon detunings. This makes Coulomb processes negligible in front of pure fermion exchanges when the detuning increases. And indeed, in essentially all nonlinear effects we have up to now studied [23-29], Coulomb interaction between excitons plays a minor role. We will see that this is also true in the present work, the dominant term in the final expression of the photoinduced current density, Eqs.(5.1,2), again being the Pauli scattering term.

Textbooks usually give the third order response to a photon field as a third order integral over time of a triple commutator [1,8,13]. This algebraic expression, although nicely compact, immediately raises a major problem since, when developed, this triple commutator makes appear the Hamiltonian acting on two-electron-hole-pair states. As the exact eigenstates of two electrons and two holes are not known except for highly simplified Hamiltonians [5,11], the exact calculation of these terms is not possible. It is however necessary to somehow control them since they contain volume linear contributions which have to be extracted in order to show that they cancel exactly, for nonlinear susceptibilities are intensive quantities. This `` cancellation problem'', which remained open for decades [2-4], has been successfully tackled recently [25,26] using the tools we have developed for interactions between composite excitons. In this framework, the calculation of the third order susceptibility turns out to be rather simple because, in it, only enter two interacting excitons, not $N$ as in many other problems we can now address. The formal elimination of these volume linear terms relies on the possibility to, in an \emph{exact way},  pass $e^{iHt}$ over an exciton creation operator. This ``trick'', which is given in Eq.(3.5), turns out to be useful not only for calculating nonlinear susceptibilities but also for any problem involving time evolution of exciton states.

Once these volume linear terms are eleminated, we are left with an expression of the nonlinear response to a photon field which reads in terms of Pauli scatterings for carrier exchanges between two excitons and Coulomb scatterings for carrier interactions, as physically reasonable since nonlinear optical effects in semiconductors are expected to come from interactions between the virtual excitons to which photons are coupled. This result was reported in ref.[26]. A somewhat more detailed version can be found in ref.[25]. However, the procedure we first used to obtain it was rather heavy: We calculated explicitly the eight terms of the triple commutator and carefully combined all the volume linear terms to show that they exactly cancel. In writing the extended version of this work, we have realized that, by calculating the first of these commutators explicitly, we are trivially left with Pauli and Coulomb terms only, so that, by doing so, we immediately get rid of all the nasty volume linear terms. It is this far simpler derivation that we here report in details. In view of its formal simplicity, it becomes easy to consider a photon field with multifrequency, as necessary to possibly describe four-wave-mixing experiments in which different laser beams would be used. In this work, we however restrict to photons all having the same circular polarization. This allows us to drop the spin indices and makes all notations far lighter. Extension to photons with different elliptical polarizations, in order to cover all possible configurations, will be given in a forthcoming paper, the goal of the present work being to concentrate on the fundamental aspects of the new procedure we propose.

Due to a bare dimensional argument, the dimensionless Pauli scatterings cannot appear in boson-exciton effective Hamiltonians [18,19] whatever the bosonization procedure to generate them is. As these Pauli scatterings control all optical nonlinear effects at large detunings, it is clear that it is not possible, through these effective Hamiltonians, to properly describe optical nonlinearities in semiconductors, nor to derive the correct expression of nonlinear susceptibilities.

Various groups [7,9,10,12], through rather different procedures, have managed to tackle the third order susceptibility while keeping the exciton composite nature exactly.  Since in the third order susceptibility, only enters the interaction of two excitons, it is not necessary to have at hand a quite general $N$-body formalism to possibly solve this two-body problem exactly. In these works, the third order response to a photon field is usually approached not through the standard textbook expression for nonlinear susceptibilities, but through its time derivative. This turns out to be quite wise as the so-called cancellation problem, which immediately appears when writing this susceptibility through a triple commutator, is then totally avoided. Comparison between the present approach and these various procedures, within the same set of notations in order to make an easy link between all of them, is clearly of interest as the obtained results can appear as rather different at first, even to a careful reader. However, this comparison, which implies to go into each of these different procedures rather in details, is definitely too long to be included here. This is why we will present it in an independent paper. 

The present paper is organized as follows. 

In section 2, we settle the problem. 

In section 3, we outline the difficulty the ``brute force'' calculation of the triple commutator raises and give the trick which allowed us to overcome it. 

In section 4, we show how we can, within a few lines, get rid of all volume linear terms and immediately reach the physically relevant Pauli and Coulomb terms. We then show how to get a compact form for these two terms.

In section 5, we discuss the final result and give its representation in Shiva diagrams.

In section 6, we conclude.

This paper also contains two appendices. In the first one, we come back to the appropriate way to describe optical nonlinearities in solid state physics, to once more stress that, although seen in many semiconductor textbooks and publications, it is not the photoinduced dipole density but the photoinduced current density which has to be used in this description. The second appendix contains a compact rederivation from scratch of the formal expression of operator mean value at third order in coupling, for completeness.

\section{The problem}

The standard way to approach optical nonlinearities is to calculate the time dependence of physically relevant operator mean values due to the wave function change induced by the coupling of the system at hand to a photon field, this field being introduced adiabatically from $t=-\infty$. As explained in appendix A, the relevant operator in solid state physics is not the dipole density $\v P(\v r)$ as used in atomic physics, but the current density $\v J(\v r)$, because wave functions for solid state systems are not spatially localized as in the case of atoms but extended over the whole sample, due to crystal periodicity. 

Since the relevant excitation operators for problems dealing with semiconductors are the exciton creation operators $B_m^\dag$, i.e., operators such that 
\begin{equation}
H_{sc}B_m^\dag|v\rangle=E_mB_m^\dag| v\rangle\ , 
\end{equation}
where $H_{sc}$ is the semiconductor Hamiltonian and $|v\rangle$ the electron-hole vacuum, the interband contribution to the current density, which is a one-electron operator, can be conveniently written in second quantization, in terms of these exciton operators,
\begin{equation}
\v J(\v r)=\sum_m\v j_m(\v r)\,B_m+\mathrm{h.c.}\ .
\end{equation}
As rederived in Appendix A, the prefactor $\v j_m(\v r)$ is related to the Kane vector $\v G$ [30] for valence-conduction transitions, given by Eq.(A.8), through $\v j_m(\v r)=-\v G\,e^{i\v Q_m.\v r}\langle\v r=\v 0|\nu_m\rangle\,L^{-D/2}$, where $\v Q_m$ is the center-of-mass momentum of exciton $m$, $\nu_m$ its relative motion index, $L$ the sample size and $D$ the space dimension.

The interaction $W_t$ between a semiconductor and a photon field can also be written in terms of exciton operators. This actually is far more convenient than using free electron-hole-pair operators as commonly done, since excitons are the physically relevant excitations in semiconductors. For a multifrequency photon field with vector potential $\v A(\v r,t)$ introduced adiabatically from $t=-\infty$ over a timescale $1/\epsilon$, 
\begin{equation}
\v A(\v r,t)=e^{\epsilon t}\sum_j\v A_j\,e^{i(\omega_jt-\v Q_j.\v r)}+\mathrm{c.c.}\ ,
\end{equation}
where $(\omega_j,\v Q_j)$ are the frequency and momentum of the photons at hand, this interaction reduces, if we only keep resonant terms (which corresponds to the so-called ``rotating wave approximation'' [31]), to
\begin{equation}
W_t=U_t+U_t^\dag\ ,
\end{equation}
where $U_t$ destroys an exciton according to
\begin{equation}
U_t=\sum_n x_n(t)\, B_n\ ,
\end{equation}
with a time-dependent weight given by
\begin{equation}
x_n(t)=e^{\epsilon t}\sum_j\mu_n^{(j)}e^{i\omega_j t}\ .
\end{equation}
The prefactor $\mu_n^{(j)}$ in $x_n(t)$ is related to the same Kane vector for valence-conduction transitions appearing in the current density, through
\begin{equation}
\mu_n^{(j)}=\v G.\v A_j\,\delta_{\v Q_n,\v Q_j}\,\langle\v r=\v 0|\nu_n\rangle L^{D/2}\ .
\end{equation}
Due to the $\langle\v r=\v 0|\nu_n\rangle$ factor, photons are only coupled to excitons with S symmetry, for their wave function to differ from zero at $\v r=\v 0$, the largest coupling being with the ground state because its wave function at the origin is the largest.

The interaction expansion of the current density mean value $\langle\v J(\v r)\rangle_t=\langle\psi_t|\v J(\v r)|\psi_t\rangle$ for a system wave function $|\psi_t\rangle$
which obeys the Schr\"{o}dinger equation $i\frac{\partial}{\partial t}|\psi_t\rangle=(H_{sc}+W_t)|\psi_t\rangle$, only has odd order terms in the semiconductor-photon interaction, since both $\v J(\v r)$ and $W_t$ create or destroy one exciton. For initial state $|\psi_{t=-\infty}\rangle$ taken as the electron-hole pair vacuum $|v\rangle$, the third order term in $W_t$ is known [1,8,13] to be given by the integral of a triple commutator (see Appendix B). Due to Eqs.(2.1) and (B.5), $\langle\v J^{(3)}(\v r)\rangle_t$ ends by reading as
\begin{equation}
\langle\v J^{(3)}(\v r)\rangle_t=\sum_m \left[j_m(\v r)\,\gamma_m(t)+j_m^\ast(\v r)\gamma_m^\ast(t)\right]\ ,
\end{equation}
where the prefactor $\gamma_m^\ast(t)$ is given by
\begin{equation}
\gamma_m^\ast(t)=(-i)^3\,\int_{-\infty}^t dt_1\,\int_{-\infty}^{t_1} dt_2\,\int_{-\infty}^{t_2} dt_3\,
S_m(t_1,t_2,t_3;t)\ .
\end{equation}
$S_m(t_1,t_2,t_3;t)$ has a nicely compact form,
\begin{equation}
S_m(t_1,t_2,t_3;t)=\langle v|\left[\left[\left[\t B_m^\dag(t),\t W(t_1)\right],\t W(t_2)\right],\t W(t_3)\right]|v\rangle\ ,
\end{equation}
which, however, is not nice at all for analytical calculation, as shown below. In the above equation,
$\t B_m^\dag(t)$ and $\t W(t)$ are the Heisenberg representations of operators $B_m^\dag$ and $W_t$, namely, $\t Z(t)=e^{iH_{sc}t}Ze^{-iH_{sc}t}$. (Note that the prefactors of $j_m^\ast(\v r)$ and $j_m(\v r)$ in Eq.(2.8) must be complex conjugate since $\langle\v J^{(3)}(\v r)\rangle_t$ is a real quantity).

The problem is to get a compact expression for $\gamma_m(t)$. In the next section, we are going to show why the ``brute force'' calculation of the triple commutator appearing in $S_m(t_1,t_2,t_3;t)$ raises a major technical problem which let it open for decades. Thanks to the many-body theory for composite-bosons we constructed [20,21], we were able to overcome this difficulty [25,26] and to derive a compact expression of $\langle\v J^{(3)}(\v r)\rangle_t$ in terms of the Pauli and Coulomb scatterings of this theory. While the expression of $\langle\v J^{(3)}(\v r)\rangle_t$ we then obtained is fully correct, it turns out that, in preparing the present manuscript, we have found a way to greatly simplify our first calculation. This more elegant derivation is given in section 4. It however is of interest to put this new calculation in its ``historical'' context by briefly presenting the ``brute force'' calculation of the triple commutator, in order to grasp the fundamental difficulty raised by a na\"{\i}ve approach to third order susceptibility.

\section{Brute force calculation}

The brute force calculation of $\gamma_m(t)$ corresponds to expand the triple commutator of Eq.(2.10) and to calculate its eight terms separately. This is actually what we first did and reported in refs. [25,26]. This expansion makes appear ``easy terms'' and ``tricky terms''. Easy terms are trivial to calculate because their intermediate state is the vacuum. Equation (2.1) leads to
\begin{eqnarray}
\langle v|\t B_{n_4}(\tau_4)\t B_{n_3}^\dag(\tau_3)\t B_{n_2}(\tau_2)\t B_{n_1}^\dag(\tau_1)|v\rangle&=&
\langle v|\t B_{n_4}(\tau_4)\t B_{n_3}^\dag(\tau_3)|v\rangle\,\langle v|\t B_{n_2}(\tau_2)\t B_{n_1}^\dag
(\tau_1)|v\rangle\nonumber\\
&=&e^{-iE_{n_4}(\tau_4-\tau_3)}\,\delta_{n_4,n_3}\,\delta_{n_2,n_1}\,e^{-iE_{n_1}(\tau_2-\tau_1)}\ .
\end{eqnarray}
In contrast, in tricky terms, the intermediate state is a two-electron-hole-pair state,
\begin{eqnarray}
\langle v|\t B_{n_4}(\tau_4)\t B_{n_3}(\tau_3)\t B_{n_2}^\dag(\tau_2)\t B_{n_1}^\dag(\tau_1)|v\rangle=
e^{-iE_{n_4}(\tau_4-\tau_3)}\,e^{-iE_{n_1}(\tau_2-\tau_1)}\nonumber\\
\times\ \langle v|B_{n_4}B_{n_3}e^{-iH_{sc}
(\tau_3-\tau_2)}B_{n_2}^\dag B_{n_1}^\dag|v\rangle\ ,
\end{eqnarray} 
so that these tricky terms are impossible to calculate exactly as the semiconductor eigenstates for two pairs are unknown.

The trouble is that the easy terms generate volume linear contributions which cannot exist in the final expression of the linear susceptibility, as this quantity is intensive. The volume linear contributions of the easy terms thus have to cancel out exactly with similar ones coming from the tricky terms. In order to prove it and extract these volume linear terms from the tricky terms, it is necessary to find a way to manipulate these tricky terms. Up to our many-body theory for composite bosons and the tools it provides, this was not possible; this is why the calculation of this triple commutator remained an open problem [2-4], except for very simple Hamiltonians for which the two-pair eigenstates could be analytically determined [5,11]. Let us now outline how the volume linear part of the tricky terms can be extracted.

In the RHS of Eq.(3.2), we can commute $e^{-iH_{sc}\tau}$ and $B_{n_2}^\dag$ by using the integral representation of the exponential, namely,
\begin{equation}
e^{-iH_{sc}t}=\int_{-\infty}^{+\infty}\frac{dx}{(-2i\pi\eta_t)}\ \frac{e^{-i(x+i\eta_t0_+)t}}{x+i\eta_t0_+-H_{sc}}
\ ,
\end{equation}
with $\eta_t=\mathrm{sign}(t)$, and the key equation for correlations with excitons [20,21], namely,
\begin{equation}
\frac{1}{a-H_{sc}}\,B_n^\dag=\left(B_n^\dag+\frac{1}{a-H_{sc}}\, V_n^\dag\right)\,\frac{1}{a-H_{sc}-E_n}\ ,
\end{equation}
$V_n^\dag$ being the creation potential for Coulomb interaction with the exciton $n$, defined as
$[H_{sc},B_n^\dag]=E_nB_n^\dag+V_n^\dag$.
This leads to (cf. ref. [21] section 6.2)
\begin{equation}
e^{-iH_{sc}t}B_n^\dag=B_n^\dag\,e^{-i(H_{sc}+E_n)t}+V_n^\dag(t)\ ,
\end{equation}
\begin{equation}
V_n^\dag(t)=\int_{-\infty}^{+\infty}\frac{dx}{(-2i\pi\eta_t)}\ \frac{e^{-i(x+i\eta_t0_+)t}}{x+i\eta_t0_+-H_{sc}}\ V_n^\dag\ \frac{1}{x+i\eta_t0_+-H_{sc}-E_n}\ ;
\end{equation}
When used in Eq.(3.2), Eq.(3.5) generates two terms. The first one, which now has $H_{sc}$ acting on one exciton, readily gives, due to Eq.(2.1),
\begin{equation}
e^{-iE_{n_1}(\tau_3-\tau_1)}e^{-iE_{n_2}(\tau_3-\tau_2)}e^{-iE_{n
_4}(\tau_4-\tau_3)}\ 
\langle v|B_{n_4}B_{n_3}B_{n_2}^\dag B_{n_1}^\dag|v\rangle\ .
\end{equation}
The above scalar product can then be expressed in terms of Pauli scatterings for carrier exchanges between excitons as [20,21]
\begin{equation}
\langle v|B_{n_4}B_{n_3}B_{n_2}^\dag B_{n_1}^\dag|v\rangle=\left(\delta_{n_4,n_1}\,\delta_{n_3,n_2}-
\lambda\left(^{n_3\ n_2}_{n_4\ n_1}\right)\right)+(n_4\leftrightarrow n_3)\ .
\end{equation}
If we then collect all the $\delta$ parts of the various tricky terms produced by this procedure, it is possible to show that they cancel out all the easy terms. We thus are left with contributions from the tricky terms coming from the $\lambda$ part of Eq.(3.8) and from the $V_n^\dag(t)$ term of Eq.(3.5). This makes the physical origin of these remaining terms quite clear: $\gamma_m(t)$ has one part coming from Pauli exclusion and one part coming from Coulomb interaction.

It is possible to calculate the Pauli contribution to $\gamma_m(t)$ analytically in terms of exciton energies and Pauli scatterings $\lambda\left(^{n_2\ n_3}_{n_1\ m}\right)$ with exciton $m$. In contrast, the fact that two-pair states play a role in this problem somehow remains, since the Coulomb part of $\gamma_m(t)$ contains scalar products like
$\langle v|B_{n_1}B_{n_2}\ \frac{1}{\omega_{j_1}+\omega_{j_2}-H_{sc}}\ V_m^\dag B_{n_3}^\dag|v\rangle$. By using one of the key equations of the composite-boson many-body theory [20,21], namely, 
\begin{equation}
[V_m^\dag,B_n^\dag]=\sum_{p,q}\xi\left(^{q\ \,n}_{p\ m}\right)B_p^\dag B_q^\dag\ , 
\end{equation}
where $\xi\left(^{q\ \,n}_{p\ m}\right)$ is the direct Coulomb scattering of this theory, we can rewrite this scalar product as
\begin{equation}
\langle v|B_{n_1}B_{n_2}\ \frac{1}{\omega_{j_1}+\omega_{j_2}-H_{sc}}\ V_m^\dag B_{n_3}^\dag|v\rangle=\sum_{n_1',n_2'}\xi\left(^{n_2'\ n_3}_{n_1'\ m}\right)\langle v|B_{n_1}B_{n_2}\,\frac{1}{\omega_{j_1}+\omega_{j_2}-H_{sc}}\,B_{n_1'}^\dag B_{n_2'}^\dag|v\rangle\ .
\end{equation}
The exact calculation of the above matrix element is not possible because it imposes the knowledge of the whole two-pair eigenstate spectrum analytically. We can however estimate such a term in two limits,

\noindent (i) close to a biexciton resonance: If $|\mathrm{XX}\rangle$ is the biexciton ground state with energy $E_\mathrm{XX}$, Eq.(3.10) then reduces to
\begin{equation}
\langle v|B_{n_1}B_{n_2}\ \frac{1}{\omega_{j_1}+\omega_{j_2}-H_{sc}}\ V_m^\dag B_{n_3}^\dag|v\rangle=\sum_{n_1',n_2'}\frac{\xi\left(^{n_2'\ n_3}_{n_1'\ m}\right)}{\omega_{j_1}+\omega_{j_2}-E_\mathrm{XX}}\,         \langle v|B_{n_1}B_{n_2}|\mathrm{XX}\rangle\langle\mathrm{XX}|B_{n'_1}^\dag B_{n'_2}^\dag|v\rangle\ ;
\end{equation}

\noindent (ii) at large detuning: Equations (3.4), (3.8) and (3.9) allow to expand Eq.(3.10) in ratios of Coulomb scattering divided by detuning which are small in the large detuning limit.

Even if the above procedure is rather straightforward once Eq.(3.5) is given, to indeed show that all the volume linear terms cancel out exactly relies on an analytical calculation which turns out to be extremely heavy. It is then reasonable  to think that if such a cancellation has to exist, there must be a clever way to show it, which avoids going through the calculation of these eight terms explicitly.

This is what we present in the next section. Due to the formal simplicity of this calculation, it becomes easy to include a multifrequency photon field, the result we first reported being obtained for one photon frequency only.

\section{Better procedure}

We come back to Eq.(2.10) and first calculate the commutator $[\t B_m^\dag(t),\t W(t_1)]$. This is easy to do thanks again to Eq.(3.5). We find that this commutator contains a Coulomb term in which enters the creation potential $V_m^\dag$, a Pauli term in which enters the deviation-from-boson operator $D_{nm}$ and a scalar. Since the scalar disappears when included in 
the triple commutator of Eq.(2.10), this readily shows that $\gamma_m(t)$, as well as all higher order susceptibilities which read in terms of higher order commutators, only have Pauli and Coulomb contributions. This result, obtained within a few lines, is already quite nice for the understanding of the physics which controls optical nonlinearities. Let us first calculate this commutator.

\subsection{Explicit calculation of the first commutator}

By using the definition of Heisenberg operators, this first commutator $[\t B_m^\dag(t),\t W(t_1)]=R_m(t_1;t)$ expands as
\begin{equation}
R_m(t_1;t)=e^{iH_{sc}t}B_m^\dag e^{-iH_{sc}(t-t_1)}W_{t_1}e^{-iH_{sc}t_1}-
e^{iH_{sc}t_1}W_{t_1} e^{-iH_{sc}(t_1-t)}B_m^\dag e^{-iH_{sc}t}\ .
\end{equation}
To calculate it, we pass $e^{-iH_{sc}(t-t_1)}$ over $B_m^\dag$ in the first term, and $e^{-iH_{sc}(t_1-t)}$ over $B_m^\dag$ in the second term. According to Eq.(3.5), this leads to split $R_m(t_1;t)$ as
\begin{equation}
R_m(t_1;t)=R_m^\mathrm{Coul}(t_1;t)+e^{iE_m(t-t_1)}e^{iH_{sc}t_1}[B_m^\dag,W_{t_1}]e^{-iH_{sc}t_1}\ .
\end{equation}
The first term is a Coulomb term as it contains the creation potential $V_m^\dag$. Its precise expression reads
\begin{equation}
R_m^\mathrm{Coul}(t_1;t)=-e^{iH_{sc}t}
V_m^\dag(t-t_1)W_{t_1}e^{-iH_{sc}t_1}e^{-iE_m(t_1-t)}-e^{iH_{sc}t_1}W_{t_1}V_m^\dag(t_1-t)
e^{-iH_{sc}t}\ .
\end{equation}
The commutator in the second term of Eq.(4.2) can be calculated by using one of the key equations for composite-boson many-body effects [20,21], namely,
$[B_{n},B_m^\dag]=\delta_{n,m}-D_{nm}$. From the definition of $W_t$ (Eqs.(2.4,5)), we find, since $[B_m^\dag,U_{t_1}^\dag]=0$, 
\begin{eqnarray}
[B_m^\dag,W_{t_1}]&=& -[\sum_{n_1}x_{n_1}(t_1)B_{n_1},B_m^\dag]\nonumber\\
&=&-x_m(t_1)+\sum_{n_1}x_{n_1}(t_1)D_{n_1m}\ .
\end{eqnarray}
By inserting this equation into Eq.(4.2), we end with 
\begin{equation}
R_m(t_1;t)=R_m^\mathrm{free}(t_1;t)+
R_m^\mathrm{Pauli}(t_1;t)+R_m^\mathrm{Coul}(t_1;t)\ ,
\end{equation}
where $R_m^\mathrm{free}(t_1;t)=-x_m(t_1)
e^{iE_m(t-t_1)}$ is a bare scalar, while $R_m^\mathrm{Pauli}(t_1;t)$, given by
\begin{equation}
R_m^\mathrm{Pauli}(t_1;t)=
e^{iH_{sc}t_1}\sum_{n_1}x_{n_1}(t_1)D_{n_1m}e^{-iH_{sc}t_1}e^{-iE_m(t_1-t)}\ ,
\end{equation}
is a Pauli contribution since it reads in terms of the deviation-from-boson operators $D_{n_1m}$.

When inserted into $\left[\left[\t B_m^\dag(t),\t W(t_1)\right],\t W(t_2)\right]$, the free part $R_m^\mathrm{free}(t_1;t)$ which is a scalar disappears; we are thus only left with Pauli and Coulomb contributions to $\gamma_m(t)$, as obtained (after a lot of efforts) through the brute force approach. Note that the same commutator $R_m(t_1;t)$ actually shows that, not only the third order susceptibility, but all higher order nonlinear susceptibilities are only made of Pauli contribution and Coulomb contribution, i.e., they only come from interactions with exciton $m$ through Pauli exclusion and Coulomb interaction, as derived from very fundamental physical arguments based on the fact that photons cannot directly interact.

\subsection{Formal structure of $\gamma_m(t)$}

Due to Eq.(4.5), $\gamma_m^\ast(t)$, splits as $\gamma_m^{\ast\mathrm{Pauli}}(t)+\gamma_m^{\ast\mathrm{Coul}}(t)$, which are obtained from Eq.(2.9) with $S_m(t_1,t_2,t_3;t)$ respectively replaced by
\begin{equation}
S_m^\mathrm{Pauli}(t_1,t_2,t_3;t)=
\langle v|\left[\left[R_m^\mathrm{Pauli}(t_1;t),\t W(t_2)\right],\t W(t_3)\right]|v\rangle
\end{equation}
and by $S_m^\mathrm{Coul}(t_1,t_2,t_3;t)$, which reads as $S_m^\mathrm{Pauli}(t_1,t_2,t_3;t)$ with $R_m^\mathrm{Pauli}(t_1;t)$ replaced by $R_m^\mathrm{Coul}(t_1;t)$.

Since $\left[\left[A_1,A_2\right],A_3\right]=(A_1A_2-A_2A_1)A_3-A_3(A_1A_2-A_2A_1)$, each of these $S_m(t_1,t_2,t_3;t)$ contains four terms. Fortunately, most of them reduce to zero. Indeed, from $D_{n'n}|v\rangle=0$ which is the key property of the deviation-from-boson operator [21], along with $D_{n'n}^\dag=D_{nn'}$, we readily see that $\langle v|R_m^\mathrm{Pauli}(t_1;t)$ and
$R_m^\mathrm{Pauli}(t_1;t)|v\rangle$ reduce to zero. Since $U_t|v\rangle=0$, we are thus left with
\begin{equation}
-S_m^\mathrm{Pauli}(t_1,t_2,t_3;t)=\langle v|U_{t_2}e^{-iH_{sc}t_2}R_m^\mathrm{Pauli}(t_1,t)e^{iH_{sc}t_3}U_{t_3}
^\dag|v\rangle\ +\ (t_2\leftrightarrow t_3)\ ,
\end{equation}
where $R_m^\mathrm{Pauli}(t_1,t)$ is the operator defined in Eq.(4.6).

We now turn to $S_m^\mathrm{Coul}(t_1,t_2,t_3;t)$ and remember [21] that the creation potential reads as $V_m^\dag=\sum B^\dag(\cdots a^\dag a+\cdots b^\dag b)$; this shows that not only $V_m^\dag|v\rangle=0$, which is the key property of the creation potential [21], but also $\langle v|V_m^\dag=0$. The state $\langle v|U_{\tau}V_m^\dag$ is also equal to zero, but for a less trivial reason: This state is a zero-pair state since $U_{\tau}$ destroys one pair while $V_m^\dag$ creates a pair. We can then insert $|v\rangle\langle v|$ in front of it to see that it reduces to zero since $V_m^\dag|v\rangle=0$. This shows that the non-zero terms of $S_m^\mathrm{Coul}(t_1,t_2,t_3;t)$ must be constructed on $\langle v|U_{\tau}U_{\tau'}V_m^\dag U_
{\tau''}^\dag|v\rangle$. By collecting them, we are left with three terms only that we split as
\begin{equation}
S_m^\mathrm{Coul}(t_1,t_2,t_3;t)=S_m^{'\mathrm{Coul}}(t_1,t_2,t_3;t)+S_m^{''\mathrm{Coul}}(t_1,t_2,t_3;t)\ ,
\end{equation}
where the first part is very similar to the two terms of $S_m^\mathrm{Pauli}(t_1,t_2,t_3;t)$,
\begin{equation}
S_m^{'\mathrm{Coul}}(t_1,t_2,t_3;t)=\langle v|U_{t_2}e^{-iH_{sc}(t_2-t_1)}U_{t_1}V_m^\dag(t_1-t)e^{iH_{sc}(t_3-t)}U_{t_3}^\dag|v\rangle\ +\ (t_2\leftrightarrow t_3)\ ,
\end{equation}
the second part reading
\begin{equation}
S_m^{''\mathrm{Coul}}(t_1,t_2,t_3;t)=-\langle v|U_{t_3}e^{-iH_{sc}(t_3-t_2)}U_{t_2}e^{-iH_{sc}(t_2-t)}V_m^\dag(t-t_1)U_{t_1}^\dag|v\rangle\,e^{-iE_m(t_1-t)}\ .
\end{equation}
To get $\gamma_m^{\ast\mathrm{Pauli}}(t)$ and $\gamma_m^{\ast\mathrm{Coul}}(t)$, we are left with calculating $S_m^\mathrm{Pauli}(t_1,t_2,t_3;t)$ and $S_m^\mathrm{Coul}(t_1,t_2,t_3;t)$ and integrating these quantities over $(t_3,t_2,t_1)$.

\subsection{Calculation of $\gamma_m^\mathrm{Pauli}(t)$}

By using $R_m^\mathrm{Pauli}$ given in Eq.(4.6), the expression of the semiconductor-photon coupling given in Eqs.(2.5,6) and Eq.(2.1), we can rewrite $S_m^\mathrm{Pauli}$, given in Eq.(4.8), as
\begin{eqnarray}
S_m^\mathrm{Pauli}(t_1,t_2,t_3;t)=-\sum_{j_1,j_2,j_3}\sum_{n_1,n_2,n_3}e^{(3\epsilon+i\omega_{j_1}+i\omega_{j_2}-i\omega_{j_3})t}\mu_{n_1}^{(j_1)}\mu_{n_2}^{(j_2)}\mu_{n_3}^{(j_3)\ast}\nonumber\\
\times\  \langle v|B_{n_2}D_{n_1m}B_{n_3}^\dag|v\rangle\,F(t_1-t,t_2-t_1,t_3-t_2)\ ,
\end{eqnarray}
where $F(\tau_1,\tau_2,\tau_3)$ reads as
\begin{equation}
F(\tau_1,\tau_2,\tau_3)=e^{(3\epsilon+i\Omega_{j_1m}+i\Omega_{j_2m}-i\Omega_{j_3m})\tau_1}e^{(2\epsilon+i\Omega_{j_2n_2}-i\Omega_{j_3n_3})\tau_2}
\left[e^{(\epsilon-i\Omega_{j_3n_3})\tau_3}+e^{\epsilon+i\Omega_{j_2n_2})\tau_3}\right]\ ,
\end{equation}
with $\Omega_{jn}=\omega_j-E_n$ being the detuning of photon $j$ with respect to exciton $n$.

The matrix element in the RHS of Eq.(4.12) readily follows from one of the key equations of the composite-boson many-body theory [20,21], namely, 
\begin{equation}
[D_{mi},B_j^\dag]=\sum_n\left[\lambda\left(
^{n\ \,j}_{m\ i}\right)+(m\leftrightarrow n)\right]B_n^\dag\ ,
\end{equation} 
where $\lambda\left(^{n\ \,j}_{m\ i}\right)$ is the Pauli scattering between two excitons. Since $\langle v|B_mB_i^\dag|v\rangle=\delta_{m,i}$, while $D_{mi}|v\rangle=0$, we get
\begin{equation}
\langle v|B_{n_2}D_{n_1m}B_{n_3}^\dag|v\rangle=\lambda\left(^{n_2\ n_3}_{n_1\ m}\right)+
(n_1\leftrightarrow n_2)\ .
\end{equation}
By inserting Eqs.(4.13,15) into Eq.(4.12), integrating over $\tau_1$, $\tau_2$, $\tau_3$ between $-\infty$ and 0 and ultimately taking the complex conjugate of the result, we find the following compact expression for $\gamma_m^\mathrm{Pauli}(t)$ in terms of Pauli scatterings between two excitons,
\begin{eqnarray}
\gamma_m^\mathrm{Pauli}(t)=-\sum_{j_1,j_2,j_3}\sum_{n_1,n_2,n_3}\frac{e^{(3\epsilon-i\omega_{j_1}-i\omega_{j_2}+i\omega_{j_3})t}
\ \ \ \mu_{n_1}^{(j_1)\ast}\mu_{n_2}^{(j_2)\ast}\mu_{n_3}^{(j_3)}}{(\Omega_{j_1m}+\Omega_{j_2m}
-\Omega_{j_3m}+3i\epsilon)(\Omega_{j_2n_2}+i\epsilon)(\Omega_{j_3n_3}-i\epsilon)}\nonumber\\
\times \left[\lambda\left(^{n_3\ n_2}_{m\ \,n_1}\right)+(n_1\leftrightarrow n_2)\right]\ .
\end{eqnarray}

\subsection{Calculation of $\gamma_m^\mathrm{Coul}(t)$}

(i) Let us start with $S_m^{'\mathrm{Coul}}$ given by Eq.(4.10). The procedure used for $S_m^\mathrm{Pauli}$ leads us to write
\begin{eqnarray}
\int_{-\infty}^{t_1}dt_2\int_{-\infty}^{t_2}dt_3\,\,S_m^{'\mathrm{Coul}}(t_1,t_2,t_3;t)=\sum_{j_1,j_2,j_3}
\sum_{n_1,n_2,n_3}\frac{\mu_{n_1}^{(j_1)}\mu_{n_2}^{(j_2)}\mu_{n_3}^{(j_3)\ast}e^{(3\epsilon+
i\omega_{j_1}+i\omega_{j_2}-i\omega_{j_3})t}}{(\epsilon+i\Omega_{j_2n_2})(\epsilon-i\Omega_{j_3n_3})}\nonumber\\
\times\ G'(t_1-t)\ ,\hspace{2cm}
\end{eqnarray}
with
\begin{equation}
G'(\tau_1)=e^{(3\epsilon+i\omega_{j_1}+i\omega_{j_2}-i\Omega_{j_3n_3})\tau_1}\,\langle v|B_{n_2}B_{n_1}
V_m^\dag(\tau_1)B_{n_3}^\dag|v\rangle\ .
\end{equation}
To go further, we use Eq.(3.6) for $V_m^\dag(\tau_1)$ and, in its last factor, replace $H_{sc}$ by $E_{n_3}$. This allows us to formally perform the $x$ integration as
\begin{equation}
V_m^\dag(\tau_1) B_{n_3}^\dag|v\rangle=\frac{e^{-iH_{sc}\tau_1}-e^{-i(E_{n_3}+E_m)\tau_1}}
{H_{sc}-E_{n_3}-E_m}\,V_m^\dag B_{n_3}^\dag|v\rangle\ .
\end{equation}
Integration of $G'(\tau_1)$ over $\tau_1$ then yields
\begin{equation}
\int_{-\infty}^0d\tau_1\,G'(\tau_1)=\frac{i}{(3\epsilon+i\Omega_{j_1m}+i\Omega_{j_2m}
-i\Omega_{j_3m})}
\langle v|B_{n_2}B_{n_1}\,T'\,V_m^\dag B_{n_3}^\dag|v\rangle\ ,
\end{equation}
where the operator $T'$ reduces to
\begin{equation}
T'=\frac{1}{3\epsilon+i\omega_{j_1}+i\omega_{j_2}-i\Omega_{j_3n_3}-iH_{sc}}\ .
\end{equation}

\noindent (ii) We now turn to $S_m^{''\mathrm{Coul}}$ defined in Eq.(4.11). We again replace the $U$'s by their expression (2.5,6), use Eq.(2.1) and integrate over $t_3$ and $t_2$. 
We find that this integral reads as Eq.(4.17), with $G'(t_1-t)$ replaced by $G''(t_1-t)$ given by
\begin{equation}
\frac{G''(\tau_1)}{(\epsilon-i\Omega_{j_3n_3})}=-\langle v|B_{n_2}B_{n_1}\,\frac{e^{(3\epsilon+i\omega_{j_1}+i\omega_{j_2}-i\omega_{j_3n_3}-iE_m-iH_{sc})\tau_1}}{2\epsilon+i\omega_{j_1}+i\omega_{j_2}-iH_{sc}}\,
V_m^\dag(-\tau_1)B_{n_3}^\dag|v\rangle\ .
\end{equation}
By inserting Eq.(4.19), with $\tau_1$ replaced by (-$\tau_1$), into the above equation, we get an expression of $G''(\tau_1)$ which again can be formally integrated over $\tau_1$. The result just
reads like Eq.(4.20) with the operator $T'$ replaced by $T''$ given by
\begin{equation}
T''=\frac{\epsilon-i\Omega_{j_3n_3}}{(3\epsilon+i\omega_{j_1}+i\omega_{j_2}-i\Omega_{j_3n_3}-iH_{sc})(2\epsilon+i\omega_{j_1}+i\omega_{j_2}-iH_{sc})}\ .
\end{equation}

\noindent (iii) The last step is to add the two contributions to $\gamma_m^{\ast\mathrm{Coul}}(t)$ coming from the triple integrals over times of $S_m^{'\mathrm{Coul}}$ and $S_m^{''\mathrm{Coul}}$. We first note that
\begin{equation}
T'+T''=\frac{1}{2\epsilon+i\omega_{j_1}+i\omega_{j_2}-iH_{sc}}\ .
\end{equation}
By using Eq.(3.10) and taking the complex conjugate of the final result, we 
end with a nicely compact form of $\gamma_m^\mathrm{Coul}(t)$ in view of the rather complicated calculation we had to perform to get it, namely
\begin{eqnarray}
\gamma_m^\mathrm{Coul}(t)=\sum_{j_1,j_2,j_3}\sum_{n_1,n_2,n_3} \frac{e^{(3\epsilon-i\omega_{j_1}-i\omega_{j_2}+i\omega_{j_3})t}\hspace{0.5cm}\mu_{n_1}^{(j_1)\ast}\mu_{n_2}^{(j_2)\ast}
\mu_{n_3}^{(j_3)}}{(\Omega_{j_1m}+\Omega_{j_2m}
-\Omega_{j_3m}+3i\epsilon)(\Omega_{j_2n_2}+i\epsilon)(\Omega_{j_3n_3}-i\epsilon)}\nonumber\\
\times\ \sum_{n_1',n_2'}\xi\left(^{n_3\ n'_2}_{m\ \,n'_1}\right)\langle v|B_{n'_1}B_{n'_2}\,\frac{1}{\omega_{j_1}+\omega_{j_2}-H_{sc}+2i\epsilon}\,B_{n_1}^\dag B_{n_2}^\dag|v\rangle\ .
\end{eqnarray}

\section{Discussion}

When compared to the Pauli contribution given in Eq.(4.16),
we see that this Coulomb contribution has a very similar structure. This leads us to write $\gamma_m(t)$ as
\begin{eqnarray}
\gamma_m(t)=e^{3\epsilon t}\sum_{j_1j_2j_3}\sum_{n_1n_2n_3} \frac{e^{-i(\omega_{j_1}+\omega_{j_2}-\omega_{j_3})t}\hspace{0.5cm}\mu_{n_1}^{(j_1)\ast}\mu_{n_2}^{(j_2)\ast}
\mu_{n_3}^{(j_3)}}{(\Omega_{j_1m}+\Omega_{j_2m}
-\Omega_{j_3m}+3i\epsilon)(\Omega_{j_2n_2}+i\epsilon)(\Omega_{j_3n_3}-i\epsilon)}\nonumber\\
\times\ \left[\Gamma_m^\mathrm{Pauli}(n_1,n_2,n_3)+\Gamma_m^\mathrm{Coul}(j_1,j_2;n_1,n_2,n_3)\right]\ ,
\end{eqnarray}
where $\Gamma_m^\mathrm{Pauli}$ and $\Gamma_m^\mathrm{Coul}$ are physically linked to interactions with the exciton $m$, through Pauli exclusion and Coulomb interaction respectively. Their precise values are given by
\begin{equation}
\Gamma_m^\mathrm{Pauli}n_1,n_2,n_3)=- \left[\lambda\left(^{n_3\ n_2}_{m\ \,n_1}\right)+(n_1\leftrightarrow n_2)\right]\ ,
\end{equation}
\begin{equation}
\Gamma_m^\mathrm{Coul}(j_1,j_2;n_1,n_2,n_3)=\sum_{n_1',n_2'}\xi\left(^{n_3\ n'_2}_{m\ \,n'_1}\right)\langle v|B_{n'_1}B_{n'_2}\,\frac{1}{\omega_{j_1}+\omega_{j_2}-H_{sc}+2i\epsilon}\,B_{n_1}^\dag B_{n_2}^\dag|v\rangle\ .
\end{equation}

\subsection{Large detuning expansion of $\gamma_m(t)$}

As previously said, the Pauli part of $\gamma_m(t)$ is fully compact since it only reads in terms of the Pauli scattering $\lambda\left(^{n_3\ n_2}_{m\ \,n_1}\right)$. In contrast, the Coulomb part cannot be obtained exactly because the full spectrum of the two-pair eigenstates is unknown. We can however note that this Coulomb term contains one additional ratio of Coulomb scattering divided by photon detuning. This makes it small far from resonance, i.e., when the concept of nonlinear susceptibility is relevant compared to absorption controlled by the Fermi golden rule, i.e., the poles of $\gamma_m(t)$.

In order to get the dominant contribution in the  large detuning limit of  $\Gamma_m^\mathrm{Coul}(j_1,j_2;n_1,n_2,n_3)$, we can make $H_{sc}$ act on the right in Eq.(5.3) and, according to Eq.(3.4), replace it  by $E_{n_1}+E_{n_2}$. By using Eq.(3.8), this leads to
\begin{equation}
\Gamma_m^\mathrm{Coul}(j_1,j_2;n_1,n_2,n_3)\simeq
\frac{\xi\left(^{n_3\ n_2}_{m\ \,n_1}\right)-\xi^\mathrm{in}\left(^{n_3\ n_2}_{m\ \,n_1}\right)+
(n_1\leftrightarrow n_2)}{\Omega_{j_1n_1}+\Omega_{j_2n_2}+2i\epsilon}\ ,
\end{equation}
where $\xi^\mathrm{in}\left(^{n_3\ n_2}_{m\ \,n_1}\right)$ is the ``in'' Coulomb exchange scattering of the composite-boson many-body theory [21]. More generally, Eqs.(3.4), (3.8) and (3.9) allow us to expand $\Gamma_m^\mathrm{Coul}(j_1,j_2;n_1,n_2,n_3)$ in ratios of Coulomb scattering divided by photon detuning.

\subsection{Shiva diagram representation}

To better grasp the physics of this third order response to a multifrequency photon field, Shiva diagrams [21,22] turn out to be once more quite valuable. 

In a linear response, one photon with frequency $\omega_{j_1}$ is coupled to the exciton $m$. This corresponds to the diagram of Fig.1(a). Since with one photon, we can only have one exciton, the linear response shown in this diagram makes appear $\mu_{m}^{(j_1)\ast}/(\omega_{j_1}-E_{m})$: The numerator $\mu_{m}^{(j_1)\ast}$ comes from the transformation of one photon with frequency 
$\omega_{j_1}$ (we here drop its $t$ dependence to simplify the notations) into the exciton with energy $E_m$, while the denominator, necessary to have a dimensionless quantity, is, as usual, the difference between the energies before and after the transition, as shown by the vertical lines in Fig.1(a); before the transition, we only have the photon $\omega_{j_1}$, and after, we have the exciton $m$.

If other photons are present, the excitons to which these photons are coupled can interact. The simplest interaction is a bare fermion exchange (see Fig.1(b)). In this diagram, we have two couplings with the ``in'' photons $\omega_{j_1}$ and $\omega_{j_2}$ which induce $\mu_{n_1}^{(j_1)\ast}$ and $\mu_{n_2}^{(j_2)\ast}$. We also have one coupling with the ``out'' photon $\omega_{j_3}$ which induces $\mu_{n_3}^{(j_3)}$. We obviously also have the Pauli scattering for fermion exchange $\lambda\left(^{n_3\ n_2}_{m\ \,n_1}\right)$, with $(n_1,n_2)$ possibly exchanged. The energy difference between the ``in'' state with photons $\omega_{j_1},\omega_{j_2}$ and the ``out'' state with exciton $m$ and photon $\omega_{j_3}$ has to appear in the denominator. Moreover, in order to have a dimensionless quantity, we need two other energy denominators since the $\mu$'s are energylike quantities. These denominators can only be the detunings associated to the excitations of the virtual excitons $n_2$ and $n_3$ which can be seen as the energy difference between states before and after these couplings (small vertical lines in Fig.1(b)). Consequently, the contribution of the diagram shown in Fig.1(b) reads 
\begin{equation}
-\frac{\mu_{n_1}^{(j_1)\ast}\mu_{n_2}^{(j_2)\ast}
\mu_{n_3}^{(j_3)}}{(\omega_{j_1}+\omega_{j_2}
-\omega_{j_3}-E_m)(\omega_{j_2}-E_{n_2})(\omega_{j_3}-E_{n_3})}\left[\lambda\left(^{n_3\ n_2}_{m\ \,n_1}\right)+(n_1\leftrightarrow n_2)\right]\ ,
\end{equation}
the minus sign being standard when the number of exchanges is odd. This exactly is the result given in Eq.(4.16).

In addition to this pure exchange, excitons can also interact through the Coulomb potential. If we restrict to one Coulomb process, we are led to diagrams (c) and (d) of Fig.1. These diagrams make appear the direct Coulomb scattering $\xi\left(^{n_3\ n_2}_{m\ \,n_1}\right)$ and the exchange Coulomb scattering $\xi^\mathrm{in}\left(^{n_3\ n_2}_{m\ \,n_1}\right)$ with a minus sign since in this last scattering, one exchange is present. However, as these scatterings are energylike quantities, we need one more energy denominator which turns out to be the energy difference between the ``in'' state with the two photons $(\omega_{j_1},\omega_{j_2})$ and the ``out'' state with the two excitons $n_1$ and $n_2$. Consequently, the contribution of diagrams (c) and (d) of Fig.1 reads
\begin{equation}
\frac{\mu_{n_1}^{(j_1)\ast}\mu_{n_2}^{(j_2)\ast}
\mu_{n_3}^{(j_3)}}{(\omega_{j_1}+\omega_{j_2}
-\omega_{j_3}-E_m)(\omega_{j_2}-E_{n_2})(\omega_{j_3}-E_{n_3})}
\ \left[\frac{\xi\left(^{n_3\ n_2}_{m\ \,n_1}\right)-\xi^\mathrm{in}\left(^{n_3\ n_2}_{m\ \,n_1}\right)+
(n_1\leftrightarrow n_2)}{\omega_{j_1}+\omega_{j_2}-E_{n_1}-E_{n_2}}\right]\ .
\end{equation}
We thus recover the expression of the large detuning limit of $\gamma_m^\mathrm{Coul}(t)$ given by Eqs.(5.1) and (5.4). 
Other Coulomb interactions between the two excitons can be included along the same line to get the detuning expansion of the third order response function. 

We see that the above discussion based on Shiva diagrams for exciton interactions [21,22] makes the various terms of $\gamma_m(t)$ rather obvious.

\subsection{Ground state exciton approximation}

From Eqs.(2.8) and (5.1), we see that $\langle\v J^{(3)}(\v r)\rangle_t$ contains a summation over excitons $m$, $n_1$, $n_2$, $n_3$, with $n=(\nu_n,\v Q_n)$ where $\nu_n$ is the relative motion index and $\v Q_n$ the center-of-mass momentum of exciton $n$.

(i) Since through the exciton-photon couplings $\mu_n^{(j)}$, relevant excitons $(n_1,n_2,n_3)$ have a momentum equal to the one of the photon to which they are coupled , while both Pauli and Coulomb scatterings conserve momentum, all exciton momenta  in Eq.(5.1) are of the order of photon momenta, i.e., small on the characteristic scale of excitons. Consequently, we can take $\v Q_m$ and all $\v Q_n$'s equal to zero.

(ii) $\v J(\v r)$  \emph{a priori} contains contributions from all excitonic levels. For practical use, it is however possible to restrict the summation to the ground state exciton: The photons at hand usually have a detuning with respect to this ground state exciton which is rather large to avoid absorption on the sides of the exciton line, but not that large to possibly mix the various exciton levels in the nonlinear susceptibility. Moreover, due to the $\langle\v r=\v 0|\nu\rangle$ factor appearing in $\v j_m(\v r)$ and in $\mu_n^{(j)}$, these couplings are the largest when $|\nu\rangle$ corresponds to the relative motion ground state $|\nu_0\rangle$. For these reasons, it is possible to replace $\nu_m$ and all $\nu_n$'s by $\nu_0$, with one exception in the Pauli term as we now show.

(iii) As a matter of fact, by looking at this Pauli term, Eq.(4.16), we see that the sum over $\nu_{n_1}$ can be readily done, which is more valuable than replacing $\nu_{n_1}$ by $\nu_0$. The only factors which depend on $\nu_{n_1}$ are the coupling $\mu_{n_1}^{(j_1)\ast}$ proportional to $\langle\nu_{n_1}|\v r=\v 0\rangle$ and the Pauli scattering $\lambda\left(^{n_3\ n_2}_{m\ \,n_1}\right)$ which, when all exciton momenta are equal to zero, reduces to [20,21]
\begin{equation}
\lambda\left(^{n_3\ n_2}_{m\ \,n_1}\right)=\sum_{\v k}\langle\nu_m|\v k\rangle\langle\nu_{n_3}|\v k\rangle\langle\v k|\nu_{n_1}\rangle\langle\v k|\nu_{n_2}\rangle\ ,
\end{equation}
where $\langle\v k|\nu_n\rangle$ is the relative motion wave function of state $n$ in momentum space. 
The closure relation $\sum_{\nu_{n_1}}|\nu_{n_1}\rangle\langle\nu_{n_1}|=I$ then allows us to readily calculate
the sum over $\nu_{n_1}$: After replacing $n_2$, $n_3$ and $m$ by $o=(\nu_0, \v Q=\v 0)$, we get
$\sum_{n_1}\mu_{n_1}^{(j_1)\ast}\lambda\left(^{o\  \,o}_{o\ n_1}\right)=\mu_o^{(j_1)\ast}\Lambda_o$,
where $\Lambda_o$ is given by
\begin{equation}
\Lambda_o=\sum_{\v k}|\langle\nu_0|\v k\rangle|^2\langle\nu_0|\v k\rangle\,\frac{\langle\v k|\v r=\v 0\rangle}{\langle\nu_0|\v r=\v 0\rangle}\ .
\end{equation}

On the other hand, in the large detuning limit of the Coulomb term given by Eqs.(5.1) and (5.4), we can replace $m$ and the $n$'s by $o=(\nu_0,\v Q=\v 0)$. Since $\xi\left(^{o\ o}_{o\ o}\right)=0$ [21], we are left with the Coulomb exchange term in which appears $\xi^\mathrm{in}\left(^{o\ o}_{o\ o}\right)$ given by [21]
\begin{equation}
\xi^\mathrm{in}\left(^{o\ o}_{o\ o}\right)=2\sum_{\v k,\v k'}V_{\v k-\v k'}|\langle\nu_0|\v k\rangle|^2\left[
|\langle\nu_0|\v k'\rangle|^2-\langle\nu_0|\v k\rangle\langle\v k'|\nu_0\rangle\right]\ .
\end{equation}

(iv) The third order current density can thus be approximated by
\begin{eqnarray}
\langle\v J^{(3)}(\v r)\rangle_t\simeq\v K_o\,e^{3\epsilon t}\sum_{j_1,j_2,j_3}
e^{-i\left[(\omega_{j_1}+\omega_{j_2}-\omega_{j_3})t-(\v Q_{j_1}+\v Q_{j_2}-\v Q_{j_3}).\v r\right]}\hspace{3cm}\nonumber\\
\times\ \frac{(\v G.A_{j_1})^\ast\,(\v G.A_{j_2})^\ast\,(\v G.A_{j_3})}{(\Omega_{j_1o}+\Omega_{j_2o}
-\Omega_{j_3o}+3i\epsilon)(\Omega_{j_2o}+i\epsilon)(\Omega_{j_3o}-i\epsilon)}\ +\ \mathrm{c.c.}\ ,
\end{eqnarray}
where the prefactor $\v K_o$ reads as
\begin{equation}
\v K_o=2\v G\left(\Lambda_o+\frac{\xi^\mathrm{in}\left(^{o\ o}_{o\ o}\right)}
{\Omega_{j_1o}+\Omega_{j_2o}+2i\epsilon}\right)\,L^D\,|\langle\v r=\v 0|\nu_0\rangle|^4\ .
\end{equation}
By replacing the ground state wave function in momentum space $\langle\v k|\nu_0\rangle$ and the Fourier transform of the Coulomb potential $V_{\v q}$ by their values in 3D or in 2D, in Eqs.(5.8) and (5.9), we can easily show that 
$\Lambda_o=\lambda_D(a_\mathrm{x}/L)^D$ and 
$\xi^\mathrm{in}\left(^{o\ o}_{o\ o}\right)=-\xi_DR_\mathrm{x}(a_\mathrm{x}/L)^D$, where $R_\mathrm{x}$ and $a_\mathrm{x}$ are the 3D exciton Rydberg and Bohr radius, the prefactors $\lambda_D$ and $\xi_D$ being equal to $7\pi/2$ and $26\pi/3$ in 3D, while they are equal to $2\pi/7$ and $(8\pi-315\pi^3/512)\simeq 6.1$ in 2D. Note that, due to the $1/L^D$ factor appearing in these scatterings, $\v K_o$ is indeed volume free.

\section{Conclusion}

This paper shows rather in details how to calculate the current density induced by a multifrequency photon field, starting from its standard textbook expression as a third order time integral of a triple commutator. It ends by a rather simple approximate expression of this current density when the ground state exciton plays a dominant role. 

The procedure which is here reported heavily relies on the tools we have developed in the many-body theory for composite excitons [20,21]. These tools mainly allow us to extract the volume linear terms appearing in the standard expression of this nonlinear susceptibility and to  show their exact cancellation. This exact cancellation, which is a necessary requirement since susceptibilities are intensive quantities, stayed an open problem for decades. The current density ends by reading in terms of Pauli scatterings for carrier exchange between two excitons in the absence of carrier interaction, and expands in terms of direct and exchange Coulomb scatterings describing Coulomb interaction between the carriers of two excitons. Once again, Shiva diagrams greatly help to understand the various contributions to the third order response.

Although the calculation of this third order susceptibility is conceptually rather trivial because two excitons only are involved, so that it is not at all a $N$-body problem, the precise algebra necessary to combine all the terms to get the volume-free final result in a compact form is not that straightforward. This is why we have found useful to report this calculation rather in details for the reader to possibly reproduce it without difficulty.

\newpage

\begin{figure}[t]
\vspace{-3cm}
\centerline{\scalebox{0.7}{\includegraphics{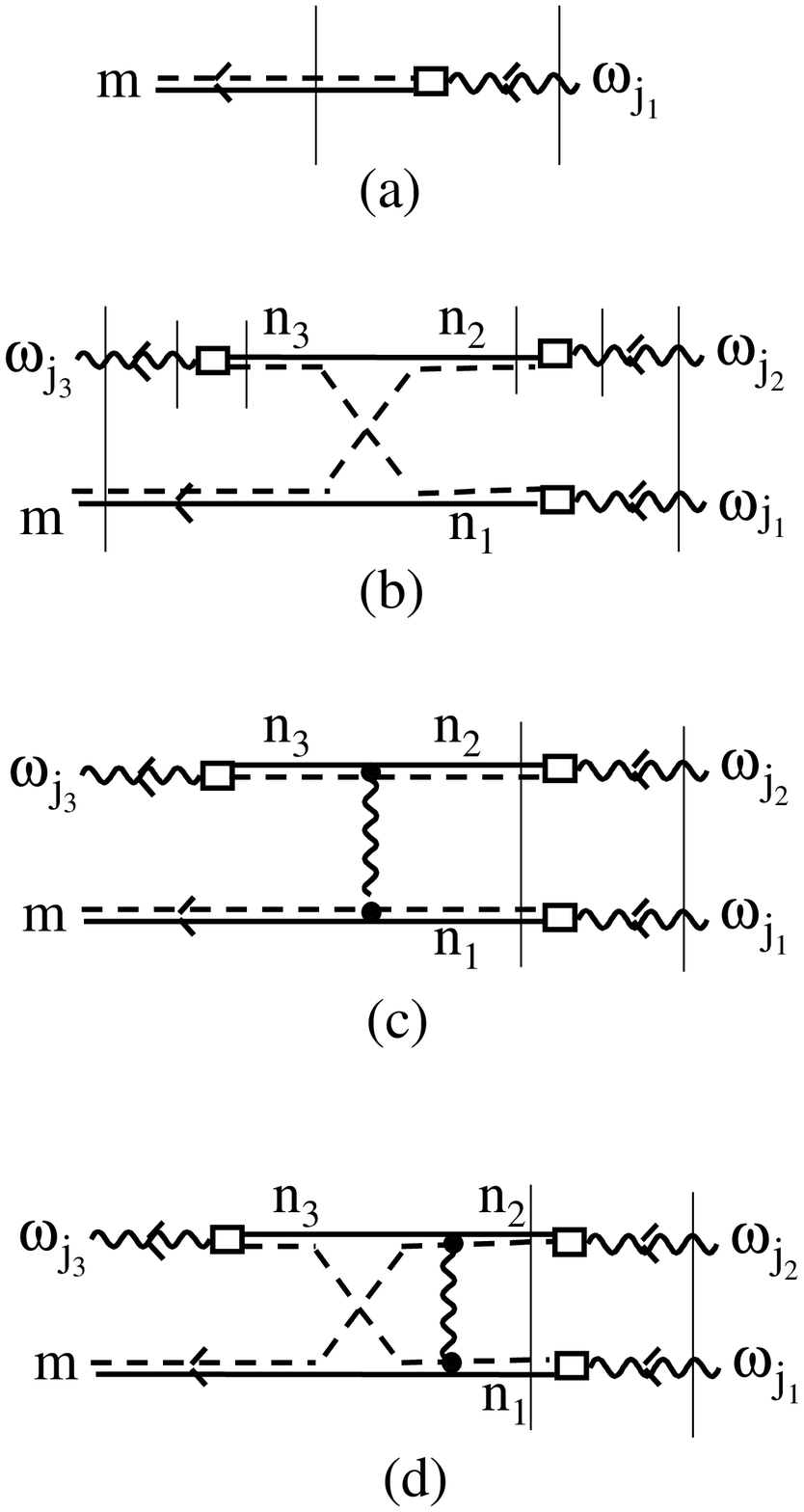}}}
\vspace{-2cm}
\caption{(a) Linear response to a photon  field: One photon with frequency $\omega_{j_1}$ transforms into one exciton in state $m$. (b) Third order response: Two photons $\omega_{j_1}$ and $\omega_{j_2}$ transform into a third photon $\omega_{j_3}$ and one exciton in state $m$. This can be done by just carrier exchange between the virtual excitons to which photons $(\omega_{j_1},\omega_{j_2})$ are coupled. (c,d) This third order response can also contain Coulomb processes. The one with one Coulomb interaction and no exchange is shown in diagram (c), while the one with one Coulomb interaction and a carrier exchange corresponds to diagram (d).}
\end{figure}

\newpage

\hbox to \hsize {\hfill \textbf{APPENDICES}
\hfill}

\vspace{0.5cm}

\appendix

\renewcommand{\theequation}{\Alph{section}.\arabic{equation}}

\section{Current density and semiconductor-photon interaction}

Although most authors describe nonlinear optical effects in terms of polarization, it is of importance to note that the physically relevant quantity for solid state physics is not the polarization, but the current density. Indeed, as explained in details for example in Cohen-Tannoudji et al.'s textbooks [32], the change from $\v A.\v p$ to $\v E.\v d$ relies on an integration by parts which is only valid for wave functions cancelling at infinity, as in the case of atomic systems, but not for solids with lattice periodicity. Since we could not find a textbook in which the following derivation is performed from scratch in a simple way, it appeared to us as useful to have it printed somewhere since the photoinduced current density is the key for any clean description of optical nonlinear effects in semiconductors.

\subsection{Current density}

The current density of a set of classical particles located at $\v r_n$ with charge $q=-|e|$ and velocities $\v v_n$, is given by
\begin{equation}
\v J(\v r)=\sum_nq\,\v v_n\,\delta(\v r-\v r_n)\ .
\end{equation}
Its Fourier transform thus reads
\begin{equation}
\v J_{\v Q}=\int d\v r\,\v J(\v r)\,e^{i\v Q.\v r}=\sum_nq\,\v v_n\,e^{i\v Q.\v r_n}\ .
\end{equation}
For quantum particles, the associated operator in first quantization is obtained, as usual, by symmetrizing the classical expression with respect to $(\v r,\v p)$; since $\v v_n=\v p_n/m$, this operator thus reads
\begin{equation}
\v J_{\v Q}=\frac{q}{2m}\sum_n\left(\v p_n\,e^{i\v Q.\v r_n}+e^{i\v Q.\v r_n}\,\v p_n\right)=\v J_{-\v Q}^\dag\ .
\end{equation}
To write it in second quantization, we use the valence and conduction Bloch states creation operators $d_n^\dag$ where $n=(c,\v k)$ or $(v,\v k)$, with $\v k$ quantized in $2\pi/L$ for a periodic solid of size $L$. The standard second quantization procedure then leads to 
\begin{equation}
\v J_{\v Q}=\sum_{n',n}\v j_{\v Q}(n',n)\,d_{n'}^\dag d_n\ ,
\end{equation}
where the prefactors are given by
\begin{equation}
j_{\v Q}(n',n)=\int d\v r\,\phi_{n'}^\ast(\v r)\frac{q}{2m}\left(\v pe^{i\v Q.\v r}+e^{i\v Q.\v r}\v p\right)\phi_n
(\v r)\ .
\end{equation}

Interband processes correspond either to $n'=(v,\v k')$ and $n=(c,\v k)$, or $n'=(c,\v k')$ and $n=(v,\v k)$. The wave function $\phi_n(\v r)$ for Bloch state $n=(c,\v k)$, with $\v r=\v R+\v{u}$ and $\v u$ restricted to a unit cell, $\v R$ running over all unit cells, reads $\phi_{c\v k}(\v r)=L^{-D/2}e^{i\v k.(\v R+\v u)}\varphi_{c\v k}(\v u)$. By using a similar form for $n'=(v,\v k')$, integration over $\v R$ in Eq.(A.5) leads to $\v k'=\v k+\v Q$. 

If we only keep these interband terms, which are the ones of interest for optical phenomena as they can lead to resonant processes with photons, we get
\begin{equation}
\v J_{\v Q}^\mathrm{inter}=\sum_{\v k}\v j_{\v Q}^{(vc)}(\v k)d_{v,\v k+\v Q}^\dag d_{c,\v k}
+\sum_{\v k}\v j_{\v Q}^{(cv)}(\v k)d_{c,\v k}^\dag d_{v,\v k-\v Q}\ .
\end{equation}
Since $\v J_{\v Q}=\v J_{-\v Q}^\dag$ due to Eq.(A.3), the prefactors in $\v J_{\v Q}^\mathrm{inter}$ must be such that $\v j_{\v Q}^{(cv)}(\v k)=\left[\v  j_{-\v Q}^{(vc)}(\v k)\right]^\ast$. We then note that (i) the $\v Q$'s of interest are usually small, (ii) the operator $\v p$ has nonzero matrix elements between valence and conduction states, since these states have different parities, the operator $\v p$ being odd. This allows us to neglect the $(\v Q,\v k)$ dependence of $\v j_{\v Q}^{(vc)}(\v k)$ and to replace it by its finite value limit   obtained for $\v Q=\v 0=\v k$. Equation (A.6) can then be replaced by
\begin{equation}
\v J_{\v Q}^\mathrm{inter}\simeq -\v G\sum_{\v k}d_{v,\v k+\v Q}^\dag d_{c,\v k}-\v G^\ast\sum_{\v k}
d_{c,\v k}^\dag d_{v,\v k-\v Q}\ ,
\end{equation}
where $\v G$, often called Kane vector [30], is given by
\begin{equation}
\v G=\frac{|e|}{m}\int_{u_c} d\v u\,\varphi_{v\v 0}(\v u)(-i\v{\nabla})\varphi_{c\v 0}(\v u)\ ,
\end{equation}
with $u_c$ being a unit cell.

The next step is to turn from valence-conduction electrons to electrons and holes according to $d_{c,\v k}^\dag=a_{\v k}
^\dag$ and $d_{v,\v k}^\dag=b_{-\v k}$. It is then appropriate to note that 
\begin{equation}
\sum_{\v k}d_{c,\v k}^\dag d_{v,\v k-\v Q}=\sum_{\v k}a_{\v k}^\dag b_{-\v k+\v Q}^\dag=
\sum_{\v p}a_{\v p+\alpha_e\v Q}^\dag b_{-\v p+\alpha_h\v Q}^\dag\ ,
\end{equation}
with $\alpha_e=1-\alpha_h=m_e/(m_e+m_h)$. By using the link between free pairs and excitons, namely,
\begin{equation}
a_{\v p+\alpha_e\v Q}^\dag b_{-\v p+\alpha_h\v Q}^\dag=\sum_{\nu}B_{\nu,\v Q}^\dag\langle\nu|\v p\rangle\ ,
\end{equation}
and by noting that $\langle\v r|\v p\rangle=e^{i\v p.\v r}L^{-D/2}$, so that 
\begin{equation}
\sum_{\v p}\langle\nu|\v p\rangle=\sum_{\v p}\langle\nu|\v p\rangle\langle\v p|\v r=\v 0\rangle L^{D/2}
=\langle\nu|\v r=\v 0\rangle L^{D/2}\ ,
\end{equation}
the sum over $\v k$ in Eq.(A.9) ends by reading in terms of exciton operators as
\begin{equation}
\sum_{\v k}d_{c,\v k}^\dag d_{v,\v k-\v Q}=\sum_m B_m^\dag \delta_{\v Q_m,\v Q}\langle\nu_m|\v r=\v 0
\rangle L^{D/2}\ .
\end{equation}
Consequently, the interband part of the current density Fourier transform $\v J_{\v Q}$ can be rewritten as
\begin{equation}
\v J_{\v Q}^\mathrm{inter}=-\v G\sum_mB_m\delta_{\v Q_m,-\v Q}\langle\v r=\v 0|\nu_m\rangle L^{D/2}-\v G^\ast\sum_mB_m^\dag\delta_{\v Q_m,\v Q}\langle\nu_m|\v r=\v 0\rangle L^{D/2}\ .
\end{equation}
Note that these two terms are not complex conjugate, in agreement with $\v J_{\v Q}=\v J_{-\v Q}^\dag$, the Fourier transform of the current density being not real, as seen from Eqs.(A.2) and (A.3).

If we now come back to the current density in $\v r$ space, its operator, given by
\begin{equation}
\v J(\v r)=\int\frac{d\v Q}{(2\pi)^D}\,e^{-i\v Q.\v r}\,\v J_{\v Q}=\frac{1}{L^D}\sum_{\v Q}e^{-i\v Q.\v r}\,
\v J_{\v Q}=\v J^\dag(\v r)\ ,
\end{equation}
has an interband contribution which ultimately reads in terms of exciton operators as
\begin{equation}
\v J^\mathrm{inter}(\v r)=\sum_m\v j_m(\v r)\,B_m+\mathrm{h.c.}\ ,
\end{equation}
in which the prefactors are given by
\begin{equation}
\v j_m(\v r)=-\v G\,e^{i\v Q_m.\v r}\,\langle\v r=\v 0|\nu_m\rangle L^{-D/2}\ .
\end{equation}
We can check that the current density has a classical expression, Eq.(A.1), which is real, so that it is represented by a Hermitian operator, in contrast with its Fourier transform $\v J_\v Q$.

\subsection{Link with semiconductor-photon interaction}

The kinetic part of a set of electrons in an electromagnetic field with vector potential $\v A(\v r,t)$ is known to be given by
\begin{equation}
\mathcal{H}_0=\sum_n \left(\v p_n-q\v A(\v r_n,t)\right)^2/2m\ ,
\end{equation}
so that the electron-photon interaction in first quantization appears as
\begin{equation}
W_t=-\frac{q}{2m}\sum_n\left(\v p_n.\v A(\v r_n,t)+\v A(\v r_n,t).\v p_n\right)\ .
\end{equation}

Since the replacement of $\v A.\v p$ by $\v E.\v d$ [32] cannot be done for the extended wave functions of solid state physics, because it relies on integration by parts with exact cancellation of the integrated term, the only correct way to describe semiconductor-photon interaction is to stay with $\v A.\v p$, as we now do. However, it can be of interest to note that the symmetry properties of $\v p$ and $\v r$ are in fact the same. It turns out that these symmetry properties are the only ones really used for most optical effects, the size of the coupling being more or less an adjustable parameter. This is why results obtained from $\v E.\v d$ instead of $\v A.\v p$ can look as correct, although derived from an incorrect first line. Nevertheless, since there is no specific advantage of working with $\v E.\v d$ instead of $\v A.\v p$, we do not see the interest of using $\v E.\v d$ in semiconductor physics, except for some readers who may find the dipole density physically more ``appealing'' than the current density.

The vector potential $\v A(\v r,t)$ for a set of photon fields with amplitude $|\v A_j|$ made of $(\omega_j,\v Q_j)$ photons, introduced adiabatically from $t=-\infty$ on the same scale $1/\epsilon$ for simplicity,  reads as
\begin{equation}
\v A(\v r,t)=e^{\epsilon t}\sum_j\left(\v A_je^{i(\omega_jt-\v Q_j.\v r)}+\mathrm{c.c.}\right)\ ,
\end{equation}
so that the electron-photon interaction splits as $W_t=\sum_jW_t^{(j)}$ , with
\begin{equation}
W_t^{(j)}=-e^{\epsilon t}e^{i\omega_jt}\v A_j.\frac{q}{2m}\sum_n\left(\v p_ne^{-i\v Q_j.\v r_n}+e^{-i\v Q_j.
\v r_n}\v p_n\right)+\mathrm{h.c.}\ .
\end{equation}
When compared to Eq.(A.3), the above equation readily shows that the electron-photon interaction is simply related to the Fourier transform of the current density through
\begin{equation}
W_t^{(j)}=-e^{\epsilon t}e^{i\omega_jt}\v A_j.J_{-\v Q_j}+\mathrm{h.c.}\ .
\end{equation}
Equation (A.12) for the interband contribution to $\v J_{\v Q}$, then gives this coupling in second quantization as
\begin{equation}
W_t^{(j)}=e^{\epsilon t}e^{i\omega_jt}\left(U_j+T_j^\dag\right)+\mathrm{h.c.}\ ,
\end{equation}
where we have set
\begin{equation}
U_j=\sum_m\mu_m^{(j)}B_m\hspace{2cm}T_j^\dag=\sum_m\tau_m^{(j)}B_m^\dag\ ,
\end{equation}
the prefactors being related to the Kane vector $\v G$ through
\begin{equation}
\mu_m^{(j)}=\v G.\v A_j\,\delta_{\v Q_m,\v Q_j}\,\langle\v r=\v 0|\nu_m\rangle\,L^{D/2}\ ,
\end{equation}
\begin{equation}
\tau_m^{(j)}=\v G^\ast.\v A_j\,\delta_{\v Q_m,-\v Q_j}\,\langle\nu_m|\v r=\v 0\rangle\,L^{D/2}\ .
\end{equation}
Among these two sets of terms, only $(U_j,U_j^\dag)$ can lead to possible resonance with photons. This is why, for photons close to the band gap, we are led to perform the so-called ``rotating wave approximation'' [31] which reduces the semiconductor-photon interaction to
\begin{equation}
W_t^{(j)}\simeq e^{\epsilon t}e^{i\omega_j t}U_j+\mathrm{h.c.}\ .
\end{equation}

\newpage

\section{Commutator expansion of operator mean value}

Let us consider an operator $A_t$, which possibly depends on time, and a state $|\psi_t\rangle$ which obeys the Shr\"{o}dinger equation,
$i\partial|\psi_t\rangle/\partial t=(H_0+W_t)|\psi_t\rangle$. We want to calculate the mean value $\langle A_t\rangle_t=\langle\psi_t|A_t|\psi_t\rangle$
as an expansion in powers of $W_t$. 

A convenient way to do it is to note that this mean value, which also reads $\langle\psi_t|e^{-iH_0t}\t A_te^{iH_0t}|\psi_t\rangle$, where $\t Z=e^{iH_0t}Ze^{-iH_0t}$ is the Heisenberg representation of operator $Z$, can be written as
\begin{equation}
\langle\psi_t|e^{-iH_0t}\sum_m|m\rangle\langle m|\t A_te^{iH_0t}|\psi_t\rangle=\sum_m\langle m|\t A_t
e^{iH_0t}|\psi_t\rangle\langle\psi_t|e^{-iH_0t}|m\rangle\ ,
\end{equation}
so that it also reads
\begin{equation}
\langle A_t\rangle_t=\mathrm{Tr}(\t A_t\,\t R_t)\ ,
\end{equation}
where $R_t=|\psi_t\rangle\langle\psi_t|$ is the projector over state $|\psi_t\rangle$.

The expansion of $\t R_t$ in powers of $\t W_t$ is easy to derive by noting that
$d/dt\, \t R_t=-i[\t W_t,\t R_t]$, so that if we expand $\t R_t$ in powers of $\t W_t$ as $\sum_{n=0}^{+\infty}
\t R_t^{(n)}$, the $\t R_t^{(n)}$'s are related by $d/dt\,\t R_t^{(n)}=-i[\t W_t,\t R_t^{(n-1)}]$.
For $\t W_{t=-\infty}=0$ and $\t R_{t=-\infty}=\t R$, this set of differential equations leads to
\begin{eqnarray}
\t R_t^{(0)}&=&\t R\ ,\nonumber\\
\t R_t^{(1)}&=&-i\int_{-\infty}^tdt_1[\t W_{t_1},\t R]\ ,\nonumber\\
\t R_t^{(2)}&=&(-i)^2\int_{-\infty}^tdt_1\int_{-\infty}^{t_1}dt_2\left[\t W_{t_1},[\t W_{t_2},\t R]\right]\ ,
\end{eqnarray}
and so on\ldots

This allows us to expand $\langle A_t\rangle_t$ as $\sum_{n=0}^{+\infty}\langle A_t\rangle_t^{(n)}$,
where $\langle A_t\rangle_t^{(n)}=\mathrm{Tr}(\t A_t\t R_t^{(n)})$. This trace is easy to write in a compact form by noting that $\mathrm{Tr}AB=\sum A_{ij}B_{ji}=\mathrm{Tr}BA$, so that $\mathrm{Tr}(AB)C=\mathrm{Tr}C(AB)$, which leads to $\mathrm{Tr}A[B,C]=\mathrm{Tr}[A,B]C$, and so on. Consequently,
\begin{eqnarray}
\langle A_t\rangle_t^{(1)}&=&-i\int_{-\infty}^tdt_1\,\mathrm{Tr}\left(\t A_t[\t W_{t_1},\t R]\right)\nonumber\\
&=&-i\int_{-\infty}^tdt_1\,\mathrm{Tr}\left([\t A_t,\t W_{t_1}]\t R\right)\ .
\end{eqnarray}
If $|\psi_{t=-\infty}\rangle=|v\rangle$, with $|v\rangle$ being $H_0$ eigenstate, we do have $|v\rangle\langle v|=
e^{iH_0t}|v\rangle\langle v|e^{-iH_0t}=\t R$, so that $\mathrm{Tr}\left(
[\t A_t,\t W_t]\t R\right)=\langle v|[\t A_t,\t W_t]|v\rangle$. A similar procedure for higher order terms  leads to write the third order term in $W_t$ as
\begin{equation}
\langle A_t\rangle_t^{(3)}=
(-i)^3\int_{-\infty}^tdt_1\int_{-\infty}^{t_1}dt_2\int_{-\infty}^{t_2}dt_3
\langle v|\left[\left[\left[\t A_t,\t W_{t_1}\right],\t W_{t_2}\right],\t W_{t_3}\right]|v\rangle\ .
\end{equation}

 \end{document}